# On the photofragmentation of $SF_2^+$: Experimental evidence for a predissociation channel

**Qun Zhang,**[1,2,a] **Rui Mao,**[1,2] **and Yang Chen**[1,2]

[1] *Hefei National Laboratory for Physical Sciences at the Microscale, University of Science and Technology of China, Hefei, Anhui 230026, People's Republic of China*

[2] *Department of Chemical Physics, University of Science and Technology of China, Hefei, Anhui 230026, People's Republic of China*

[a] Author to whom all correspondence should be addressed.  Electronic mail: qunzh@ustc.edu.cn

We report on the first observation of the photofragmentation dynamics of $SF_2^+$. With the aid of state-of-the-art *ab initio* calculations on the low-lying excited cationic states of $SF_2^+$ performed by Lee *et al.* [J. Chem. Phys. **125**, 104304 (2006)], a predissociation channel of $SF_2^+$ is evidenced by means of resonance-enhanced multilphoton ionization spectroscopy.  This work represents a second experimental investigation on the low-lying excited cationic states of $SF_2^+$.  [The first one is the He I photoelectron spectrum of $SF_2^+$ reported by de Leeuw *et al.* three decades ago, see Chem. Phys. **34**, 287 (1978).]   © *2008 American Institute of Physics.*

In light of its potential importance in the semiconductor industry, $SF_2$ reactive intermediate has recently received increased attention from both experimentalists[1,2] and computational chemists.[3]  When compared with its neutral counterpart, the cationic $SF_2^+$ species is, however, still far from being well characterized. Even from the computational chemistry side, $SF_2^+$ has witnessed quite meager investigations.[4-7] More scarce remain the spectroscopic data and dynamics of $SF_2^+$. The only experimental study available for the low-lying excited states of $SF_2^+$ comes from the He I photoelectron (PE) spectrum, which was reported by de Leeuw *et al.* three decades ago.[8] The spectral features observed in the PE spectrum were assigned to ionizations from the $\tilde{X}^1A_1$ ground state of $SF_2$ to the $\tilde{X}^2B_1$ ground state and five low-lying excited states ($\tilde{A}^2A_1$, $\tilde{B}^2B_2$, $\tilde{C}^2B_2$, $\tilde{D}^2A_1$, and $\tilde{E}^2A_2$) of $SF_2^+$.[8] A recent *ab initio* study employing the high-level RCCSD(T) method was carried out by Lee *et al.*[6] to calculate the potential energy functions (PEFs) for the $\tilde{X}^1A_1$ state of $SF_2$ and for the six states of $SF_2^+$ listed above. Since the experimental research regarding $SF_2^+$ has lagged far behind, any useful information on the low-lying states of $SF_2^+$ from the experimental side is highly desirable.[6] In this Letter, we report the first observation on the photofragmentation dynamics of $SF_2^+$ by recording the resonance-enhanced multilphoton ionization (REMPI) spectra of $SF_2$ and its concomitant $SF^+$ fragment, from which a predissociation channel occurring in the low-lying excited states of $SF_2^+$ is revealed to account for the observed fragmentation.

In the present experiment, $SF_2$ molecules in their ground state were produced by the pulsed dc discharge in a pulsed molecular beam of $SF_6$ (30% in Ar). Details of the experimental apparatus and procedure can be found in our previous publications.[9,2] The upper part of Figure 1 shows the (2 + 1) REMPI excitation spectrum of $^{32}SF_2$ ($m/z$ = 70) in the laser wavelength range of 293 − 325 nm. Based on previous work,[9,10] we can readily assign this spectrum to the $\widetilde{E}(4p) \leftarrow \leftarrow \widetilde{X}^1A_1$ and $\widetilde{F}(4p) \leftarrow \leftarrow \widetilde{X}^1A_1$ transitions of $SF_2$. [See Ref. 9 for details regarding the assignments given in Fig. 1.] The lower part of Fig.1 shows the simultaneously recorded REMPI "daughter" spectrum of $SF^+$ ($m/z = 51$). Notably, it is only from a certain onset wavelength of ~310.9 nm that this "daughter" spectrum resembles its "parent" one. Such type of spectral resemblance usually implies that a mechanism of resonance-enhanced multiphoton ionization followed by dissociation (REMPID) takes effect.[2,9-11]

To facilitate the following discussions, we put all relevant elements in an energy level diagram (Figure 2). The horizontal dotted lines in Fig. 2 denote the adiabatic ionization energies (AIEs) for the six cationic states.[8,6] The three-photon energies used in our experiment lie $>$ 1.5 eV below the cationic dissociation limit $[SF^+(X^3\Sigma^-) + F(^2P^o)]$, hence it is natural to consider only the four-photon excitation, *i.e.*, a (2 + 2) REMPID scheme is most likely to account for the observed fragmentation. Care is then taken in two four-photon energy regions, indicated by the lower crossed window (~15.37 − 15.95 eV, four photons of ~322.7 − 310.9 nm) and the upper hatched window (~15.95 − 16.81 eV, four photons of ~310.9 − 295.0 nm) in Fig. 2. The observed $SF^+$ fragment spectrum (Fig. 1) clearly indicates that the four-photon excitation into the lower window which lies between the AIEs of the $\widetilde{D}^2A_1$ and $\widetilde{E}^2A_2$ states does not yield any $SF^+$ fragment ions, while that into the upper window which transcends the AIE of the $\widetilde{E}^2A_2$ state does, as shown in Fig. 2.

Thanks to the RCCSD(T) PEFs data reported by Lee *et al.*,[6] we are able to construct the potential energy curves along the S−F symmetric stretching for the $\tilde{X}^1A_1$ state of SF$_2$ as well as for the $\tilde{X}^2B_1$, $\tilde{C}^2B_2$, $\tilde{D}^2A_1$, and $\tilde{E}^2A_2$ states of SF$_2^+$. The constructed curves in the S−F bond length range of 1.1 − 1.85 Å are displayed in Figure 3. The reasons that we take into account only the S−F symmetric stretching mode lie in that (1) our REMPI spectra do not show any identifiable vibrational structure associated with the asymmetric stretching mode, and (2) the excitation of the bending mode, though partly observed in our spectra (such as the $2_0^1$, $1_0^1 2_0^1$, and $1_0^2 2_0^1$ bands of the $\tilde{F}$ (4p) ← ← $\tilde{X}^1A_1$ transition of SF$_2$, see Fig. 1), does not contribute to the S−F bond cleavage.

Four photons of the onset wavelength (~310.9 nm) can bring the neutral precursor SF$_2$ from its ground state (about the equilibrium bond length of ~1.59 Å,[12,6] in terms of a vertical transition) to an energy position of ~15.95 eV, in the vicinity of which the $\tilde{D}^2A_1$ state crosses the $\tilde{C}^2B_2$ and $\tilde{E}^2A_2$ states at ~1.61 Å and ~1.74 Å (see Fig. 3), respectively. However, interactions and avoided crossing between these states are not possible because of different symmetry. Moreover, no direct dissociation will occur via these states since they all are not dissociative.[6] In order to reasonably accommodate our experimental observation, we therefore anticipate that a predissociation channel might be open right above the onset energy (~15.95 eV), *i.e.*, there may exist another hitherto unidentified cationic state of repulsive nature. At ~15.95 eV, such a repulsive state, if it exists, may cross the $\tilde{D}^2A_1$ state at ~1.71 Å or cross the $\tilde{C}^2B_2$ state at ~1.85 Å (indicated by the two red circles in Fig. 3). It is noteworthy that the products $SF^+(X^3\Sigma^-) + F(^2P^o)$ could combine to form a cationic state of $SF_2^+$ with only three possible symmetries: $^2A_2$, $^2B_1$, or $^2B_2$, among which only the state of $^2B_2$ can couple to the $\widetilde{C}\,^2B_2$ state to yield predissociation owing to avoided crossing.

To summarize, we have presented the firm experimental evidence for a predissociation channel occurring in the low-lying cationic excited states of $SF_2^+$. Nevertheless, confirmation of our above predictions calls for further investigations.

The authors acknowledge the NSFC for financial support, and thank J. Shu, X. Zhou, Q. Li, S. Yu, X. Ma, S. Tian, and Y. Bai for helpful discussions. Q. Z. also thanks the University of Science and Technology of China for support grants.

[1] M. E. Jacox, J. Phys. Chem. Ref. Data **32**, 1 (2003); and references therein.

[2] Q. Zhang, X. G. Zhou, Q X. Li, S. Q. Yu, and X. X. Ma, J. Chem. Phys. **128**, 144306 (2008).

[3] see, *e.g.*, Y. J. Liu and M. B. Huang, Chem. Phys. Lett. **360**, 400 (2002); K. L. Baluja and J. A. Tossell, J. Phys. B **37**, 609 (2004); J. Czernek and O. Živný, Chem. Phys. Lett. **435**, 29 (2007); and references therein.

[4] K. K. Irikura, J. Chem. Phys. **102**, 5357 (1995).

[5] Y. S. Cheung, Y. J. Chen, C. Y. Ng, S. W. Chiu, and W. K. Li, J. Am. Chem. Soc. **117**, 9725 (1995).

[6] E. P. F. Lee, D. K. W. Mok, F. T. Chau, and J. M. Dyke, J. Chem. Phys. **125**, 104304 (2006).

[7] J. Czernek and O. Živný, Chem. Phys. **344**, 142 (2008).

[8] D. M. de Leeuw, R. Mooyman, and C. A. de Lange, Chem. Phys. **34**, 287 (1978).

[9] Q. X. Li, J. N. Shu, Q. Zhang, S. Q. Yu, L. M. Zhang, C. X. Chen, and X. X. Ma, J. Phys. Chem. A **102**, 7233 (1998).

[10] R. D. Johnson III and J.W. Hudgens, J. Phys. Chem. **94**, 3273 (1990).

[11] X. G. Zhou, Q. X. Li, Q. Zhang, S. Q. Yu, Y. Su, C. X. Chen, and X. X. Ma, J. Elec. Spec. Relat. Phenom. **108**, 135 (2000).

[12] D. R. Johnson and F. X. Powell, Science **164**, 950 (1969).

[13] E. R. Fisher, B. L. Kickel, and P. B. Armentrout, J. Chem. Phys. **97**, 4859 (1992).

# FIGURE CAPTIONS

FIG. 1.   (Color online) (2 + 1) REMPI spectrum of $^{32}SF_2$ ($m/z$ = 70) (upper trace) and the fragment REMPI spectrum of $^{32}SF^+$ ($m/z$ = 51) (lower trace) simultaneously recorded in the laser wavelength range of 293 − 325 nm.

FIG. 2.   (Color online) Energy level diagram that is related to the REMPID process discussed in the text.   [a] from Ref. 8;   [b] from Ref. 13.

FIG. 3.   (Color online) Potential energy curves for the $\tilde{X}^1A_1$ state of $SF_2$ and for the $\tilde{X}^2B_1$, $\tilde{C}^2B_2$, $\tilde{D}^2A_1$, and $\tilde{E}^2A_2$ states of $SF_2^+$, all of which are constructed using the calculated RCCSD(T) PEFs data reported in Ref. 6.

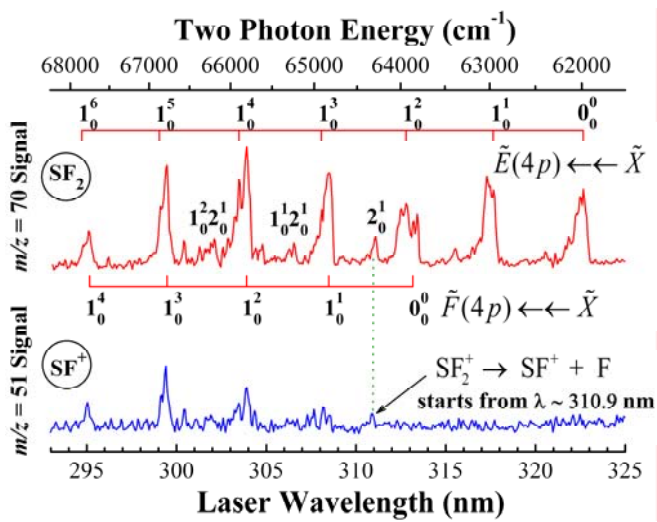
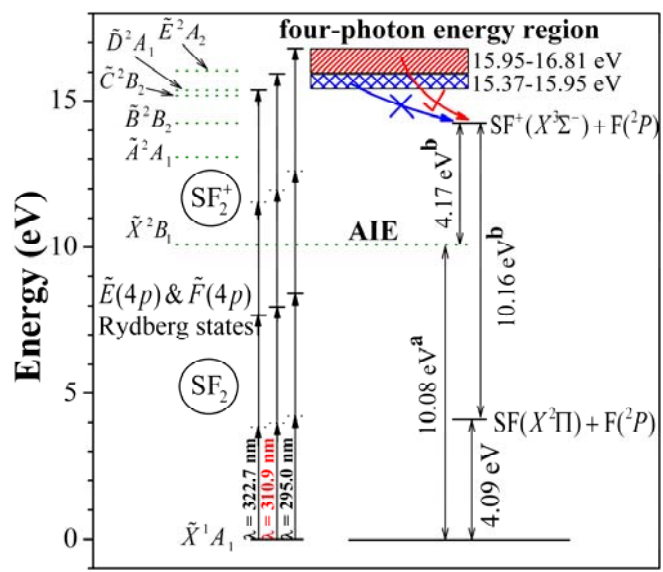

**Fig. 1**  **Fig. 2**

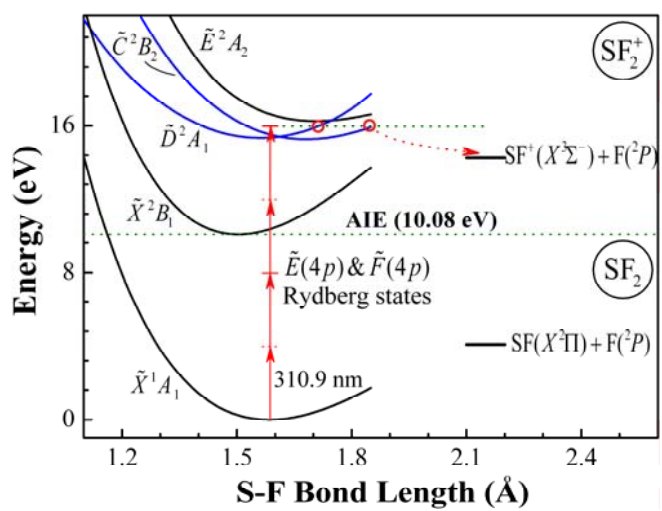

**Fig. 3**